\documentclass[12pt,preprint]{aastex}

\shorttitle{Observations of Molecular Lines toward G28.20-0.04N}
\shortauthors{Qin et al.}

\begin{document}


\title{HIGH-RESOLUTION OBSERVATIONS OF MOLECULAR LINES TOWARD THE HOT CORE
G28.20-0.04N }


\author{Sheng-Li Qin\altaffilmark{1}, Maohai Huang\altaffilmark{1},
Yuefang Wu\altaffilmark{2}, Rui Xue\altaffilmark{1}, Sheng
Chen\altaffilmark{1}}

\altaffiltext{1}{National Astronomical Observatories, Chinese
Academy of Sciences, Beijing, 100012, China; slqin@bao.ac.cn,
mhuang@bao.ac.cn, jerryxue@gmail.com, schen@bao.ac.cn.}

\altaffiltext{2}{Department of Astronomy, Peking University,
Beijing, 100871, China; yfwu@vega.bac.pku.edu.cn.}


\begin{abstract}
We present the results from arcsecond resolution observations of
various line transitions at 1.3 mm toward hypercompact HII region
G28.20-0.04N. With the SMA data, we have detected and mapped the
transitions in the CH$_{3}$CN, CO, $^{13}$CO, SO$_{2}$, OCS, and
CH$_{3}$OH molecular lines as well as the radio recombination line
H30$\alpha$. The observations and analysis indicate a hot core
associated with G28.20-0.04N. The outflow and possible rotation
are detected in this region.

\end{abstract}


\keywords{HII regions --- ISM:individual (G28.20-0.04N) ---
ISM:kinematics and dynamics --- ISM:molecules --- radio lines: ISM
--- star:formation}


\section{INTRODUCTION}
The scenario of massive star formation remains unclear and has
been observationally challenging because of large distances,
clustered formation environments, and shorter evolutionary
timescales of massive stars. Ultracompact HII (UCHII) regions are
considered signposts of massive star formation, but do not
represent the earliest stage of massive star forming process
(e.g., Churchwell 2002). Hot molecular cores are defined as
compact ($\leq$ 0.1 pc), dense ($\geq 10^7$ cm$^{-3}$) and warm
($\geq$ 100 K) molecular cloud cores (Kurtz et al. 2000). The
observations from various wavelengths have suggested that hot
cores are the sites for massive star formation, and that they
represent the early phase of the evolution prior to UCHII regions
(Kurtz et al. 2000; Gibb et al. 2000, 2004; Churchwell 2002). So
far only twenty or so hot cores from high-resolution observations
have been reported in the literature. The observations of hot
cores are important for understanding the evolutionary sequence
and physical conditions of massive star formation. Hypercompact
HII (HCHII) regions have smaller sizes and higher densities
compared with UCHII regions, and they probably correspond to a
transition phase from hot cores to UCHII regions.

The HCHII region G28.20-0.04N at a distance of 5.7 kpc has been
observed at radio and millimeter wavelengths (Fish et al. 2003;
Sollins et al. 2005; Sewilo et al. 2004, 2008; Keto et al. 2008).
Masers, rotation, inflow, and outflow in G28.20-0.04N have been
revealed from observations of molecular and radio recombination
lines at centimeter wavelengths (Argon et al. 2000; Menten 1991;
Sollins et al. 2005; Sewilo et al. 2008). However, the star
formation environment and physical conditions of the hot core in
this region are still poorly understood.

Due to the compact and dense nature of hot cores, high angular
resolution observations using molecular lines with high critical
densities and excitation temperatures at (sub)millimeter
wavelengths are crucial to uncovering the physical conditions and
kinematics in G28.24-0.04N. In this letter, we present the
Submillimeter Array (SMA) \footnote {The Submillimeter Array is a
joint project between the Smithsonian Astrophysical Observatory
and the Academia Sinica Institute of Astronomy and Astrophysics
and is funded by the Smithsonian Institution and the Academia
Sinica.} observations of molecular lines at 1.3 mm toward the hot
core in G28.24-0.04N.

\section{DATA}
The data are from SMA archive. Observations toward G28.20-0.04N
were carried out with the SMA in September 2005, at 220 GHz (lower
sideband) and 230 GHz (upper sideband) with a frequency resolution
of 0.4125 MHz and an angular resolution of 1$^{\prime\prime}$. The
mean system temperature was 100 K. In the database, QSO 3c454.3
and Uranus were observed for bandpass and flux-density
calibrations. QSOs 1743-038 and 1911-201 were also observed for
the antenna-based gain corrections. The calibration and imaging
were done in Miriad\footnote
{http://sma-www.cfa.harvard.edu/miriadWWW}. Multiple lines were
detected in both sidebands. Using the task UVLIN in Miriad, the
continuum level was determined by a linear fitting to the
line-free channels. Then, the calibrated (u, v) data were
separated for two output (u, v) data sets: the continuum and the
continuum-subtracted spectral lines. In order to improve
sensitivity, we smoothed the H30$\alpha$, CO and $^{13}$CO lines
to 2 km~s$^{-1}$ velocity resolution, and smoothed rest of the
lines to 1 km~s$^{-1}$ velocity resolution. Self-calibration was
performed to the continuum data. The gain solutions from the
continuum were applied to the line data. The synthesized beam size
of the continuum and line images with robust weighting was
approximately 1.3$^{\prime\prime}\times 0.8^{\prime\prime}$
(P.A.=$-89.6^{\circ}$).

\section{RESULTS}

The molecular lines were identified using the CDMS and JPL
databases (M\"{u}ller et al. 2005; Pickett et al. 1998), and the
data by Sutton et al. (1985) \& Nummelin et al. (1998). The
molecular lines CH$_{3}$CN, CO, $^{13}$CO, SO$_{2}$, OCS,
CH$_{3}$OH, and the radio recombination line H30$\alpha$ were
detected and identified from the lower and upper sidebands of the
SMA observations. Table 1 lists the parameters of the detected
molecular transitions.

Combining the lower and upper sideband data, we made continuum
image from line-free channels at 1.3 mm, as shown in Fig. 1(a).
The rms (1 $\sigma$) noise level is 0.003 Jy beam$^{-1}$. The
observations show an unresolved core peaked at
R.A.(J2000.0)=18$^{\rm h}$42$^{\rm m}$58.120$^{\rm s}$ ($\Delta
{\rm R.A.}=\pm 0.02^{\prime\prime}$),
decl.(J2000.0)=$-4^{\circ}$13$^{\prime}$57.40$^{\prime\prime}$
($\Delta {\rm decl.}=\pm 0.02^{\prime\prime}$) with a peak
intensity of 0.67$\pm$0.02 Jy beam$^{-1}$. The total flux density
and deconvolved source size from Gaussian fit are approximately
0.89$\pm$0.03 Jy and 0.6$^{\prime\prime}\times0.2^{\prime\prime}
({\rm P.A.}=-10.1^{\circ})$, respectively. The thick contours in
Fig. 1(a) are the integrated intensity map of the radio
recombination line H30$\alpha$. The SMA observations show that the
peak position of the continuum is consistent (within
$0.03^{\prime\prime}$) with that of the H30$\alpha$ line image in
G28.24-0.04N. Most likely the continuum emission in this
observation comes from free-free emission (see Fig. 10 of Keto et
al. 2008).

The line images were constructed from the maps of the
continuum-subtracted spectral channels. Figure 1(b) shows the
integrated intensity contours of the CH$_{3}$CN (12$_{2}-11_{2}$)
transition. The major axis of the line image is approximately
1.4$^{\prime\prime}$, corresponding to a projected linear size of
$<$0.04 pc. In order to avoid the cancellation of the integrated
intensities of the emission (positive values) and absorption
(negative values) in the overlapping region along the
line-of-sight, the emission and absorption in the CO and $^{13}$CO
were handled separately for the integrated intensity maps, as done
by Qin et al. (2008). Figure 1(c) presents the integrated
intensity map of the $^{13}$CO. The absorption against the
continuum and the emission were observed as shown in Fig. 1(c). No
absorption was evident in the H30$\alpha$ and CH$_{3}$CN lines.
Away from the continuum, the absorption was not observed. Figure 2
presents the spectra of the molecular line transitions. The
spectra were extracted from the channel maps at the peak positions
of the line images. In Fig. 2, the emission spectra have similar
profiles, showing single emission peak with similar line center
velocity and line width. Gaussian fits were performed on all the
spectra. The peak intensity ($I_{p}$), full width at half maximum
($\Delta V$), and central line velocity ($V_{LSR}$) from the
Gaussian fits are summarized in Table 1. The radial velocities
range from 95 to 96 km~s$^{-1}$ (except for $^{13}$CO, CO and
H30$\alpha$ lines). Compared with the SO$_{2}$, OCS, CH$_{3}$OH
and CH$_{3}$CN, the $^{13}$CO line with lower excitation
temperature shows both absorption and emission components.
Absorption can be observed if the excitation temperature of gas in
front of the background continuum is lower than the brightness
temperature of the continuum. Absorption is also observed in the
CO line.

\section {DISCUSSIONS}
\subsection{\emph{Hot Core}}
Methyl cyanide (CH$_{3}$CN) has been proved as an ideal probe to
determine the kinetic temperature and column density of molecular
gas (e.g., Remijan et al. 2004). Seven K-components of the
CH$_3$CN ($J=12-11$) transition were detected by the SMA
observations. From the integrated intensities, in the limits of
optically thin and local thermodynamic equilibrium (LTE), the
rotation temperature and column density are estimated using a
rotation temperature diagram (Goldsmith \& Langer 1999; Liu et al.
2002). Figure 3 shows the rotation temperature diagram. A linear
least-square fit is performed toward the seven CH$_{3}$CN
transitions. The derived rotation temperature and beam-averaged
column density are 308$\pm$22 K and (1.6$\pm$0.3)$\times$10$^{16}$
cm$^{-2}$, respectively. In the Orion molecular cloud, the
CH$_{3}$CN transitions were only detected in the hot core and the
compact ridge, giving fractional abundances relative to H$_{2}$ of
7.8$\times10^{-9}$ and 3.2$\times10^{-10}$, respectively (Blake et
al. 1987). Taking source size of 0.04 pc and the fractional
abundance in the Orion cloud, we inferred the H$_{2}$ density of
$>10^{7}$ cm$^{-3}$ in G28.24-0.04N. The relatively higher gas
temperature (308$\pm$22 K), H$_{2}$ density ($>10^{7}$ cm$^{-3}$)
and smaller size ($<$0.04 pc) indicate a \emph{hot core} in this
region.

Following Gibb et al. (2000), we adopt $3 \times 10^{24}$
cm$^{-2}$ as a typical H$_2$ column density. Using this value, we
calculate a fractional abundance of CH$_3$CN relative to H$_2$ of
$5 \times 10^{-9}$. The derived abundance is close to that in the
Orion hot core (Blake et al. 1987). The derived rotation
temperature agrees (within 2$\sigma$ of the least-square fit) with
the value of 280 K estimated by the NH$_{3}$ lines (Soillns et al.
2005). The similar gas temperatures derived from the CH$_{3}$CN
and NH$_{3}$ lines suggest a close relationship between the two
N-containing species. The gas-phase chemistry likely dominates the
pathways to produce CH$_{3}$CN (Rodgers \& Charnley 2001), in
which NH3 evaporates from grain surfaces and CH3CN is then formed
via gas-phase reactions at high temperatures. Based on the high
temperature (300 K) gas-phase chemical model (Rodgers \& Charnley
2001), if NH$_{3}$ is injected for the chemical reactions, the
derived fractional abundance of the CH$_{3}$CN of
5$\times$10$^{-9}$ corresponds to a time-scale of 1.5$\times
10^{4}$ yr which is in good agreement with those
(1.9$\times10^{3}-5.7\times10^{4}$ yr) observed in other hot cores
(Kurtz et al. 2000). The CH$_{3}$CN, SO$_{2}$, OCS and CH$_{3}$OH
spectra have similar radial velocities and line widths, suggesting
that the N-bearing, O-bearing and S-bearing molecules probably
originated from the hot core.
\subsection{\emph{Kinematics}}
For the $^{13}$CO absorption component in Fig. 1(c), the derived
optical depth of the line peak is 0.57 by use of the line to
continuum ratio (cf. equation (1) of Qin et al. 2008). The
intensity of 1 Jy~beam$^{-1}$ in this observation corresponds to a
brightness temperature of 26 K under the Rayleigh-Jeans
approximation. Given the beam filling factors of the line and
continuum of 0.5, the upper limit of the excitation temperature in
the $^{13}$CO line is 35 K (cf. equation (5) of Qin et al. 2008).
Assuming the gas is in LTE and the emission component (with
integrated flux density of 9.1 Jy~beam$^{-1}$ km~s$^{-1}$) is
optically thin with abundances of [CO]/[$^{13}$CO]=89 and
[CO]/[H$_{2}$]=1$\times10^{-4}$, the H$_{2}$ column densities of
the absorption and emission components are $\sim$
4.6$\times10^{22}$ and 6.8$\times10^{22}$ cm$^{-2}$, respectively.
The major axes of the absorption and emission components are
approximately 0.025 and 0.04 pc. The estimated H$_2$ densities and
masses are 5.9$\times10^{5}$ cm$^{-3}$, 1.5 M$_{\odot}$ and
5.5$\times10^{5}$ cm$^{-3}$, 1.9 M$_{\odot}$ for the absorption
and emission components, respectively. Compared with the rotation
temperature and H$_{2}$ density traced by CH$_{3}$CN lines, the
lower excitation temperature and H$_{2}$ densities suggest that
the $^{13}$CO is located outside of the hot molecular core. The
average $V_{LSR}$ from multiple lines is an approximation of the
systemic velocity of the molecular cloud (Sutton et al. 1991). The
average $V_{LSR}$ of the emission lines (except for H30$\alpha$,
CO and $^{13}$CO) is 95.4$\pm$0.5 km~s$^{-1}$ (see Table 1). We
adopt 95.4$\pm$0.5 km~s$^{-1}$ as the systemic velocity of the
cloud. The spectrum in the $^{13}$CO line (see Fig. 2(j)) shows
that the absorption is blue-shifted and the emission is
red-shifted with respect to the systemic velocity, consisting of a
P Cygni-like profile. A NE-SW velocity gradient and broad line
wings in NW-SE direction have been observed in the radio
recombination lines (Sewilo et al. 2008). The authors proposed a
rotating motion responsible for the observed velocity gradient and
an outflow along the rotational axis in the NW-SE direction. In
Fig. 1(c), the peaks of the absorption and emission are aligned
with the rotational axis. The spectral profile along with the
$^{13}$CO distribution appear to indicate an outflow from the
molecular core, where the blue-shifted absorption (in front of the
continuum) is flowing toward us and the red-shifted emission
(behind the continuum) is moving away in the opposite direction.
The CO spectrum shows a complex with red-shifted emission and both
blue and red-shifted absorption with respect to the systemic
velocity (see Fig. 2(i)). The infall component in this region were
observed by red-shifted absorption in the NH$_{3}$ line (Sollins
et al. 2005). Likely the blue-shifted outflow traced by the CO
line in this observation is mixed with the infall. Higher spatial
and spectral resolution observations are needed to verify the
kinematics traced by the CO line.

The position-velocity diagram across the peak of the continuum
along the NE-SW direction is constructed from the OCS line (see
Fig. 4). In Fig. 4, the velocities of the two emission peaks are
95 and 97 km~s$^{-1}$ with 0.2$^{\prime\prime}$ separation,
respectively, indicating a velocity gradient in NE-SW direction.
Assuming a rotating motion along NE-SW with a rotation axis in the
NW-SE direction (Sewilo et al. 2008) and an equilibrium between
rotational and gravitational forces, the dynamical mass
responsible for the rotation can be estimated by $M=V^{2}\times
r/G$, where $V$ and $r$ are the velocity difference and spatial
separation of the emission peaks, respectively; $G$ is the
gravitational constant. The derived dynamical mass is 25
M$_{\odot}$, which is consistent with the mass derived by Sewilo
et al. (2008).

\acknowledgments

We thank the SMA staff for making the observations possible. We
thank the anonymous referee for constructive comments on the
paper. Y.F. Wu acknowledges the support from NSFC under grant
10733030.











\clearpage
\begin{deluxetable}{llccccccccc}
\tabletypesize{\scriptsize} \tablenum{1} \tablewidth{0pt}
\tablecaption{Molecular Line Parameters} \tablehead{
\colhead{Molecule }&
                     &
\colhead{Transition} & \colhead{Frequency} & \colhead{$E_{u}$} &
\colhead{$I_{P}$}&
\colhead{$\Delta V$}& \colhead{$V_{LSR}$}&\colhead{Channel rms}&\\
&&&(MHz)&(K)&
  (Jy~beam$^{-1}$)
& (km~s$^{-1}$)&(km~s$^{-1}$) &(Jy~beam$^{-1}$) } \startdata

CO$^{a}$      &&2--1&230538.00 &17   & --0.2$\pm$0.07 & 9.9$\pm$3.0&79.9$\pm$1.3 &0.07 \\
        && \dots   & \dots  &  \dots   &--0.3$\pm$0.07 &10.3$\pm$4.5&94.1$\pm$1.6& \dots  \\
        && \dots   & \dots         &  \dots   & 0.28$\pm$0.09 & 7.1$\pm$2.1&103.1$\pm$1.3 & \dots  \\
$^{13}$CO$^{a}$ &&2--1&220398.68 &16   & --0.29$\pm$0.05 & 4.8$\pm$0.8&77.1$\pm$0.4 &0.05 \\
        && \dots   & \dots         &  \dots   & 0.3$\pm$0.05 & 6.3$\pm$0.9&104.3$\pm$0.4 & \dots  \\
OCS  &&19--18 &231060.98 &111  & 1.1$\pm$0.06 & 3.7$\pm$0.2&95.4$\pm$0.1 &0.06 \\
SO$_{2}$&& 11$_{1,11}$--10$_{0,10}$&221965.21 &60&1.5$\pm$0.07 &4.1$\pm$0.2 & 95.7$\pm$0.1 &0.07\\
CH$_{3}$OH A&& 10$_{2}$--9$_{3}$&231281.10 &166&0.3$\pm$0.07 &4.6$\pm$0.8 & 95.5$\pm$0.3 &0.07\\
CH$_{3}$CN && 12$_{0}$--11$_{0}$$^{b}$&220747.26 &69&1.1$\pm$0.08 &3.8$\pm$0.3 & 95.1$\pm$0.1 &0.08\\
          && 12$_{1}$--11$_{1}$$^{b}$&220743.01 &76&1.2$\pm$0.08 &4.9$\pm$0.4 & 95.0$\pm$0.1 &0.08\\
          && 12$_{2}$--11$_{2}$&220730.26 &97&0.9$\pm$0.08 &4.1$\pm$0.3 & 95.3$\pm$0.1 &0.08\\
          && 12$_{3}$--11$_{3}$&220709.02 &133&1.0$\pm$0.08 &4.2$\pm$0.3 & 95.4$\pm$0.1 & 0.08\\
          && 12$_{4}$--11$_{4}$&220679.29 &183&0.6$\pm$0.08 &4.8$\pm$0.5 & 95.4$\pm$0.2 &0.08\\
          && 12$_{5}$--11$_{5}$&220641.09 &247&0.7$\pm$0.07 &3.9$\pm$0.3 & 95.3$\pm$0.1 &0.07\\
          && 12$_{6}$--11$_{6}$&220594.43 &325&0.6$\pm$0.08 &3.3$\pm$0.4 & 95.8$\pm$0.2 &0.08\\
H30$\alpha$$^{c}$&& \dots&231900.96   & \dots&0.9$\pm$0.2 &20.9$\pm$0.6&92.5$\pm$0.2&0.06\\
\enddata

\tablenotetext{\it a} {Negative values of the intensity ($I_{P}$)
in the CO and $^{13}$CO spectra indicate the absorption. Positive
values indicate the emission.}

\tablenotetext{\it b} {The two CH$_{3}$CN transitions
(12$_{0}$--11$_{0}$) and (12$_{1}$--11$_{1}$) were blended.}

\tablenotetext{\it c}{H30$\alpha$ line parameters are cited from
the paper by Keto et al. 2008.}

\end{deluxetable}
\clearpage

\begin{figure}
\epsscale{0.4}\plotone{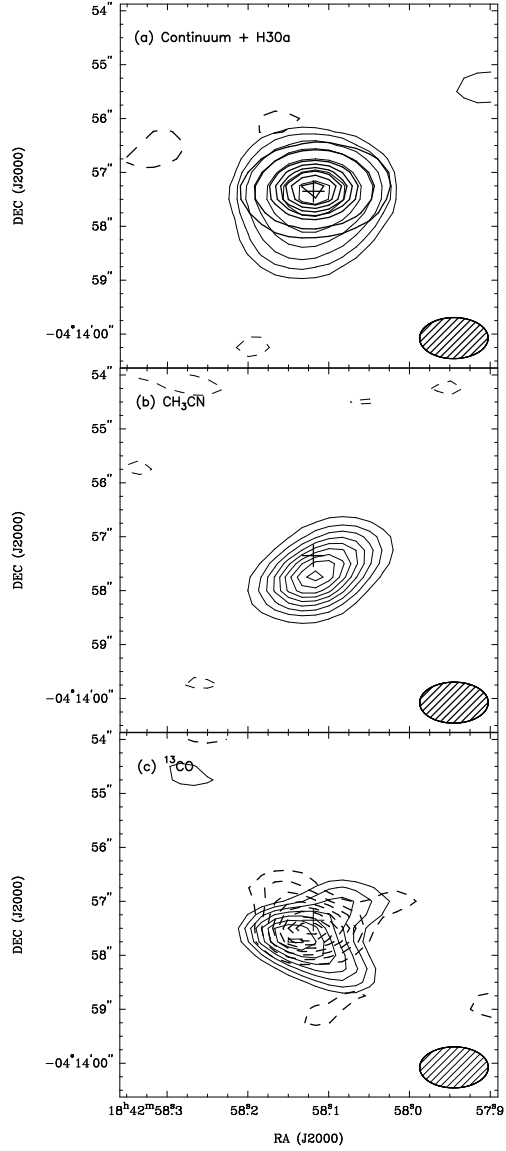}
\caption{Continuum and spectral images. (a): The continuum (thin
contours) is superimposed on the H30$\alpha$ radio recombination
line image (thick contours). The synthesized beam is
1.3$^{\prime\prime}\times 0.8^{\prime\prime}$,
P.A.=$-89.6^{\circ}$ (lower-right corner). The rms (1 $\sigma$)
noise level of the continuum is 0.003 Jy~beam$^{-1}$. The contours
are $-$4, 4, 8, 16, 32, 64, 128, 190, and 220 $\sigma$. The cross
symbol indicates the peak position of the continuum source. The
contours of the H30$\alpha$ integrated intensity are $-$1.24,
1.24, 3.68, 6.13, 8.58, 11.03, 13.48, 15.93, 18.38, 20.83, and
23.28 Jy~beam$^{-1}$ km~s$^{-1}$; (b): The CH$_{3}$CN
(12$_{2}-11_{2}$) integrated intensity contours. The levels are
$-$0.74, 0.74, 1.23, 1.72, 2.21, 2.70, 3.19, 3.68, 4.17, and 4.66
Jy~beam$^{-1}$ km~s$^{-1}$; (c): The $^{13}$CO (2-1) integrated
intensity contours. The levels are $-2, -1.8, -1.6, -1.4, -1.2,
-1, -0.8, -0.6$, 1, 1.3, 1.6, 1.9, 2.2, 2.5, 2.8, 3.1
Jy~beam$^{-1}$ km~s$^{-1}$; the positive and negative values
indicate the emission and absorption, respectively.}
\end{figure}
\clearpage

\begin{figure}
\epsscale{0.7}
\plotone{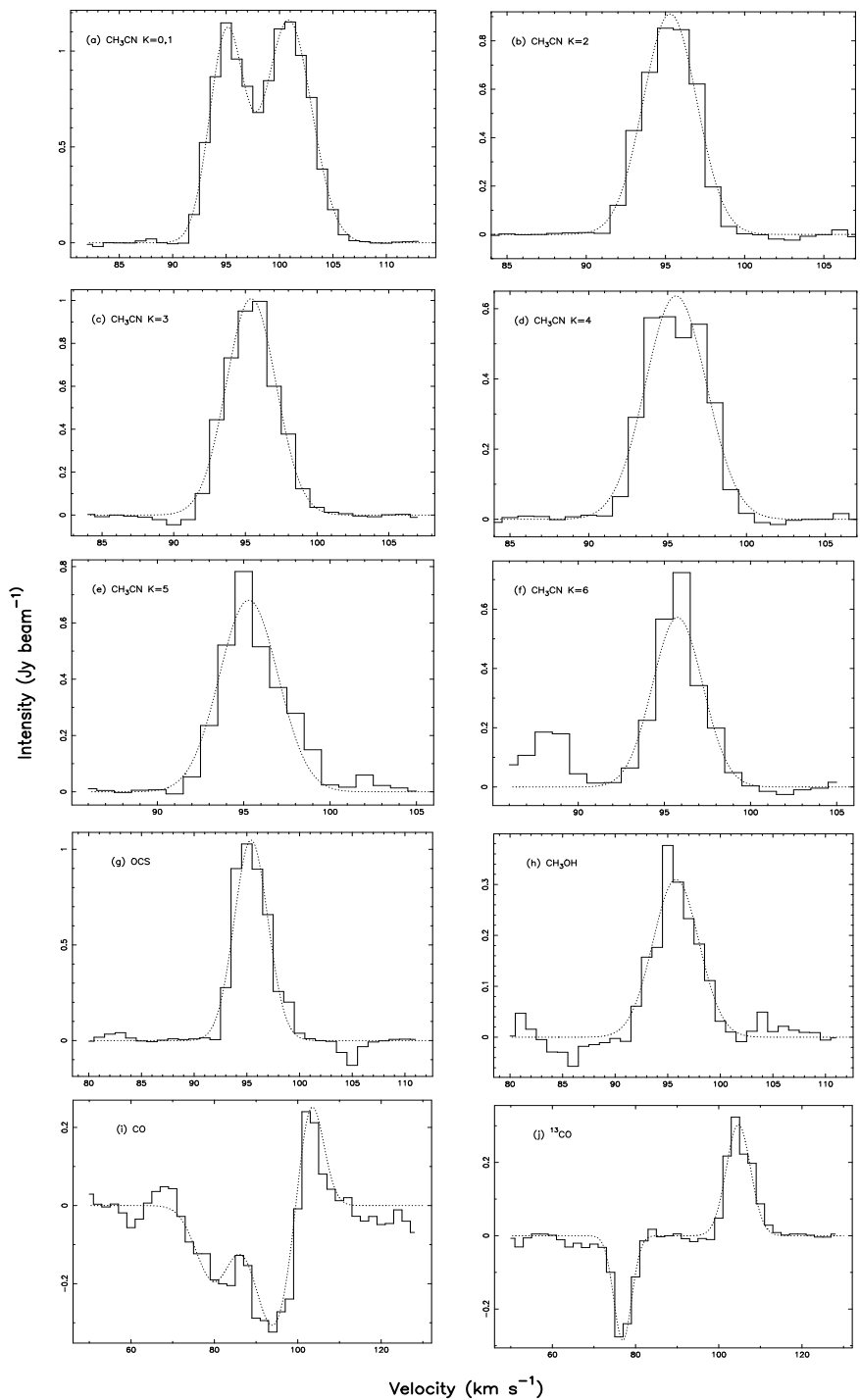}\caption{Molecular spectra averaged
over one beam. The solid and dashed curves are the observed
spectra and the Gaussian fitting to the spectra, respectively.
(a), (b), (c), (d), (e) and (f) are for the various $K$ components
in the CH$_{3}$CN (12$-$11); (g), (h), (i) and (j) are for  OCS
(19$-$18), CH$_{3}$OH (10$_{2}-9_{3}$), CO (2$-$1) and $^{13}$CO
(2$-$1) transitions, respectively.  The spectra are Hanning
smoothed for better signal to noise ratios. Negative and positive
values of the intensity in the CO and $^{13}$CO spectrum indicate
absorption and emission, respectively.}
\end{figure}

\clearpage

\begin{figure}
\epsscale{1.0} \plotone{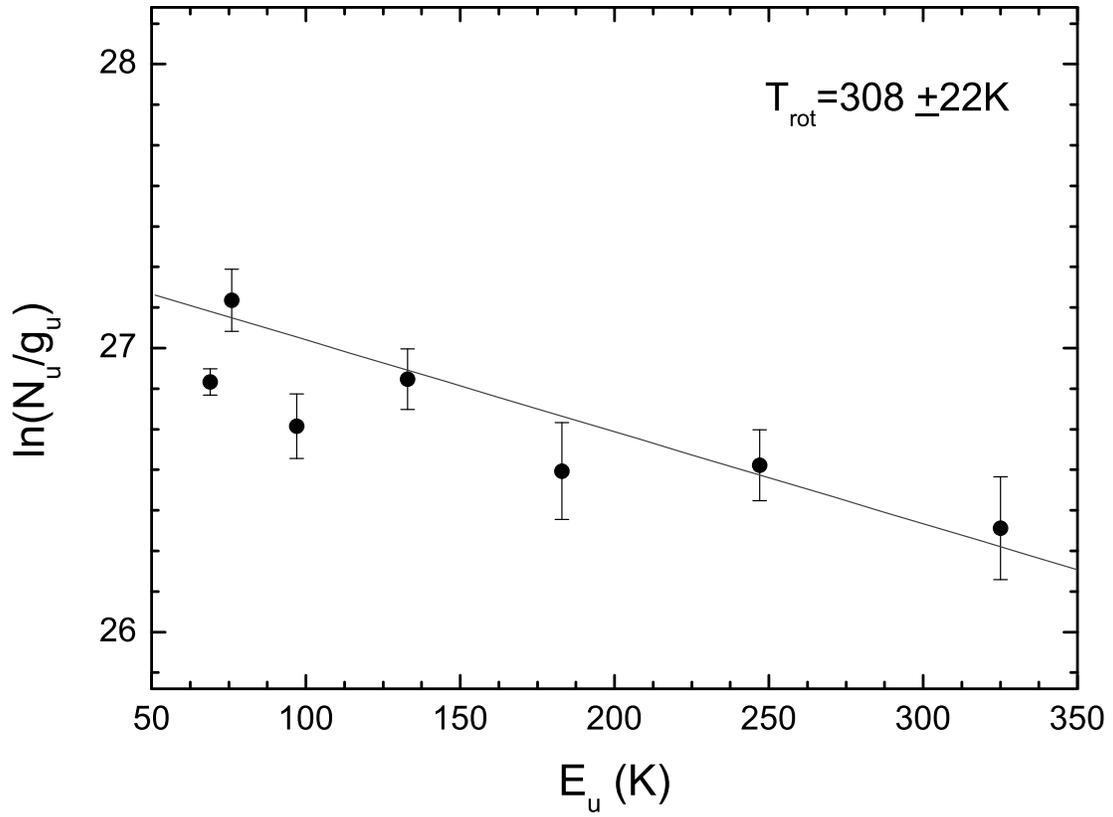} \vspace{1.5cm} \caption{Rotation
temperature diagram for the observed CH$_{3}$CN transitions. The
linear least-square fit (indicated by solid line) gives a rotation
temperature of 308$\pm$22K. The vertical bars indicate the errors
of ln (N$_{\rm u}$/g$_{\rm u}$) transferred from the integrated
intensities. }

\end{figure}

\begin{figure}
\epsscale{0.8} \plotone{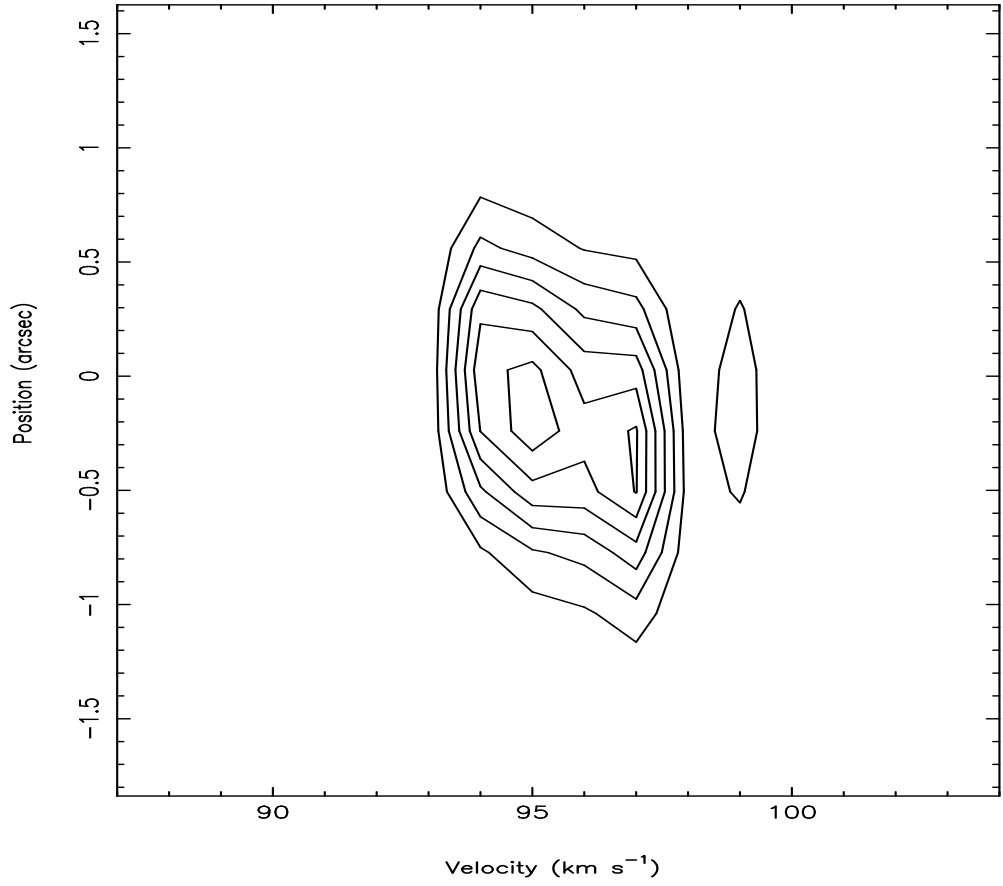} \vspace{1.5cm}
\caption{Position-velocity diagram cutting along NE-SW direction
across the continuum peak. The diagram is constructed from the OCS
transition. The contours are 0.19, 0.38, 0.56, 0.75, 0.94 and 1.13
Jy~beam$^{-1}$. }

\end{figure}


\end{document}